\documentclass[12pt,a4paper,english,superscriptaddress,aps,nofootinbib]{revtex4}
\usepackage{amsmath,amssymb,graphicx}
\makeatletter
\usepackage{babel}
\usepackage[active]{srcltx}
\usepackage{graphicx,color}
\usepackage{changebar}
\usepackage{hyperref}
\hypersetup{
	colorlinks=true,
	urlcolor= blue,
	citecolor=blue,
	linkcolor= blue,
	bookmarks=true,
	bookmarksopen=false,}

\usepackage[T1]{fontenc}
\usepackage{ae}
\usepackage[ansinew]{inputenc}
\usepackage{amsmath}
\usepackage{braket}
\usepackage{mathtools}
\usepackage{slashed}
\usepackage{empheq}
\usepackage{tikz}
\usepackage{multirow}
\usetikzlibrary{decorations.pathmorphing}
\usepackage[caption = false]{subfig}

\usepackage{graphicx}
\usepackage{amsfonts}
\usepackage{amsmath}
\newcommand{\be}{\begin{equation}}
	\newcommand{\ee}{\end{equation}}
\newcommand{\bea}{\begin{eqnarray}}
	\newcommand{\eea}{\end{eqnarray}}

\begin{document}
	\title{Topological structures in a non-canonical perturbative dynamics of a cuscuton-like model}
	
\author{F. C. E. Lima}
\email[]{E-mail: cleiton.estevao@fisica.ufc.br}
\affiliation{Universidade Federal do Cear\'{a} (UFC), Departamento do F\'{i}sica - Campus do Pici, Fortaleza, CE, C. P. 6030, 60455-760, Brazil.}

\author{A. Yu. Petrov}
\email[]{E-mail: petrov@fisica.ufpb.br}
\affiliation{Departamento de F\'{\i}sica, Universidade Federal da 
		Para\'{\i}ba,\\
		C. P. 5008, 58051-970, Jo\~ao Pessoa, Para\'{\i}ba, Brazil}
	
\author{C. A. S. Almeida}
\email[]{E-mail: carlos@fisica.ufc.br}
\affiliation{Universidade Federal do Cear\'{a} (UFC), Departamento do F\'{i}sica - Campus do Pici, Fortaleza, CE, C. P. 6030, 60455-760, Brazil.}	


\begin{abstract}
In this work,  a possible description for quantum dynamics of the cuscuton within the sigma-model approach is presented. Lower order perturbative corrections and the structure of divergences are found. Motivated by the results generated by the perturbative approach, we investigate the existence of topological structures in the cuscuton-like model. The structures we study are, first, kink-like configurations in two dimensional spacetime,  second, vortex solutions in three dimensional one with gauge field ruled by the Maxwell term. In fact, to show the existence of kink solutions it is needed to introduce a standard dynamics term in the cuscuton-like model. Then, a numerical approach (interpolation method) is used and the solution of the scalar field is presented. On the other hand, for the study of topological vortices, we reorganized the energy density to obtain, for convenience, equations similar to those canonical vortex structures, namely, the Maxwell-Higgs model. In fact, even for this particular case, we observed the existence of structures with localized energy and quantized magnetic flux in a given topological sector. We also show that when the model does not spontaneously break the symmetry, the $(2+1)D$ model only admits the so-called non-topological field solutions.
\end{abstract}

\maketitle
\thispagestyle{empty}

\section{Introduction}

Topological structures are present in various scenarios of our world. We find topological structures when we analyze the behavior of a wave propagating in the sea \cite{rajaraman}, or when we study superconductivity phenomena \cite{Machida,Bernevig} or even when we study cosmological objects \cite{nielsen,Ren,Brihaye,Goldberger,Cvetic,Cvetic1,DeWolfe}. Theoretically the topological defects arise as a consequence of a theory that has a spontaneous breaking of symmetry \cite{Giamarchi}. In fact, the simplest structures found in the literature are: the kink \cite{Vachaspati}, the vortices \cite{Vachaspati, Manton} and the monopoles \cite{Manton}. It is important to mention that each class of solution is associated with the spatial dimension of the model. Therefore, kink appears when we study $(1+1)$D models, vortices appear in $(2+1)$D models and monopoles arise in four-dimensional ones. 

In recent years several researchers have turned their attention to the study of these structures \cite{Adam, Adam1, Cart, Queiruga}, in particular, kinks and vortices \cite{Vachaspati}. Part of this interest is motivated by the fact that such structures discussed in field theory have direct applications in condensed matter physics \cite{Fradkin}, since they are similar. In principle, if we consider a qualitative point of view, such topological structures are formed during a phase transition and are related to the breaking of some symmetry of the model \cite{Giamarchi}. In this way, we can particularly relate the vortex structures that arise in a context of field theory to Abrikosov's vortices known as characteristic phenomena in condensed matter physics \cite{Abrik,Abrik1}. 

As a motivation for the study of these structures we have some applications such as: in the study of vortices in topological superconductors \cite{Bernevig}, the study of topological structures in two-dimensional quantum gravity \cite{Verlinde}, topological solutions describing the multivortex dynamics in Abelian theories \cite{Chae}, structures in models with $k$-defects \cite{Babichev}, study of generalized models \cite{Bazeia,Lima,Lima1} and non-generalized ones \cite{Kim,Lee,Lee1,Jackiw}.

It is also important to mention that topological structures appear in non-canonical models \cite{Andrade}. The motivation for the study of such configurations arises from theories such as the inflationary evolution \cite{Armen1}, where it is possible that generalized non-canonical terms in the absence of an interaction lead to an inflationary evolution \cite{Armen2,Armen3}. In fact, we have the supposition that some generalized non-canonical models can give explanations about the accelerated evolution of the universe. It is important to note that in the topological defects context, non-canonical models can provide new behaviors, as news configurations of fields, generating new classes of solutions and structures. 

In this work, we are interested in the study of topological structures in a non-canonical model known as cuscuton model whose first example was originally introduced by Afshordi \textit{et. al.} \cite{cus}. However, it is worth noting that all the theories presented so far discuss the non-canonical models only within the cosmological scenario or extra dimensions context \cite{cuscos,1,2}. In contrast to this, we found for the first time in the literature a possible description for the quantum dynamics of cuscuton basing on the approach developed for the nonlinear sigma model (see an example of this approach in Ref. \cite{GN}). Motivated by the results of the perturbative approach, we built cuscuton-like models in $(1+1)$ and $(2+1)$ dimensional spacetime, and investigated the existence of topological structures that each model admits. We show that for a particular class of solutions the vortex structures of the model have a localized energy and a quantized magnetic flux in a given topological sector. 

Our work is organized as follows: in Sec. II, we consider the cuscuton-like models in $(1+1)$ and $(2+1)$ dimensional spacetime and investigate the existence of solutions of topological structures of the model for certain forms of the potential. In Sec. III, we develop the perturbative approach and, using the auxiliary field methodology similar to that applied in the context of the nonlinear sigma model, we construct a possible description for the quantum dynamics of the cuscuton. In Sec. IV, we summarize our results and present some conclusions.

\section{Topological structures in Cuscuton-like model}

In this section, we investigate the possible emergence of topological structures in 2-dimensional and 3-dimensional models with the interaction term derived from the perturbative dynamics of the cuscuton model.

It is important to say that the cuscuton model appears as an alternative or a new model of dark energy, which, although generally non-uniform, lacks a degree of dynamic freedom \cite{cuscos}. The word ``Cuscuton''  has an origin in Latin derived from a name of a parasitic planta called cuscuta. It is applied in this context due to the fact that this model is a new type of restriction system that allows a new dynamic class, namely, non-canonical models.

In this work, our motivation is twofold. First, we seek a direct description of the quantum dynamics of cuscuton based on the nonlinear sigma model through of perturbative method (see Sec. III). Second, we want to study how topological structures can appear in cuscuton-like models. However, it is observed that the usual cuscuton dynamics only produces non-topological structures. To get around this point, we use the correction term of the perturbative theory and build a new cuscuton-like model. In this model, the interaction spontaneously breaks the symmetry and guarantees the existence of topological structures. The non-polynomial correction term that appears in quantum theory is interesting due to its ability to amplify the magnetic flux intensity and energy of vortex structures, which can be a useful property for its experimental detection \cite{LA1}.

\subsection{The cuscuton-like model in (1+1) dimensional spacetime (2D)}

\subsubsection{Basic definitions}

Usually, topological effects are studied in canonical models interacting with Higgs polynomial potentials \cite{rajaraman,nielsen}. We want now to investigate the possible existence of topological structures in the cuscuton-like model. In order to analyze whether the model supports kink-like topological solutions, that is, concentrating on studies in $2D$, let us consider the following action:
\begin{align}
	\label{cus}
    S=\int\, d^2x\, [\mu^2\sqrt{\partial_\mu\phi\partial^{\mu}\phi}-V(\phi)].
\end{align}

In our theory, we can firstly choose, as an example, a particular potential of the form
\begin{align}
	\label{part}
    V=\frac{\lambda}{4!}\phi^4+\frac{\lambda^2\phi^4}{256\pi^2}\bigg[\text{ln}\bigg(\frac{\phi^2}{M^2}\bigg)-\frac{25}{6}\bigg].
\end{align}

This potential is inspired by perturbative calculations in the usual $\phi^4$ scalar field model \cite{CW}, but in principle, our study will not be conceptually modified if we treat other potentials. It is interesting to note that logarithmic potentials are widely considered within various studies of effective models, see f.e. our previous paper \cite{Lima} and references therein. As we see below, this potential allows to obtain nontrivial topological structures, namely, kinks and vortices. At the same time, the calculations of one-loop corrections in the cuscuton model itself are detailed in the next section. Actually, we can assume that the cuscuton, being effectively of the first order in derivatives, ``dominates'' over the usual $\partial_{\mu}\phi\partial^{\mu}\phi$ kinetic term in the infrared domain of the derivative expansion, so, the theory (\ref{cus}) can be treated as the infrared limit of the theory being a sum of the usual one-loop corrected $\phi^4$ theory and the cuscuton model. Here,  
\begin{align}
 \label{pot}
    V_1\simeq\frac{\lambda^2\phi^4}{256\pi^2}\bigg[\text{ln}\bigg(\frac{\phi^2}{M^2}\bigg)-\frac{25}{6}\bigg],
\end{align}
 is the one-loop contribution \cite{CW}.
 
In fact, only the finite terms are taken into account when we consider topological structures in the model,  once its potential behaves as a combination of the classical $\phi^4$ theory term and quantum corrections. Here we assume that the divergent parts of any contribution  do not affect the class of solutions of the topological structures since they are cancelled by the corresponding counterterms.
 
From the point of view of non-perturbative studies, nothing prohibits the use of interaction (\ref{pot}). Based on the arguments that support the study of topological structures, the logarithmic interaction can be considered as long as the gauge invariance is preserved and the model admits the existence of a spontaneous symmetry breaking. In fact, the interaction (\ref{pot}) has these properties preserved. Therefore, the study of topological structures with the logarithmic interaction is possible. Here it is interesting to mention that the spontaneous symmetry breaking of the model occurs when the values of the parameters $M$ and $\lambda$ are correctly adjusted. In this direction, see Ref. \cite{Belen} where Belendryasova {\it et. al.} brings a discussion of the study of topological structures with logarithmic nonlinear interaction. 

Knowing that the logarithmic interaction is the contribution responsible for inducing the emergence of topological structures in the theory, and that the contribution of $\phi^4$ term  (with a single vacuum) is not able to produce topological structures, we consider the potential (\ref{pot}), which is reminiscent of the perturbative calculus of Refs. \cite{Lima,CW} for study of the topological solutions in our work.

For the study of topological structures it is necessary to investigate the model's Euler-Lagrange equation (\ref{cus}). In this way, we obtain that the equation of motion for static field configurations is given by 
\begin{align}
\label{kink}
    \frac{\lambda}{64\pi^2}\bigg\{\lambda\bigg[\text{ln}\bigg(\frac{\phi^2}{M^2}\bigg)-\frac{25}{6}\bigg]+\frac{1}{2}\bigg\}\phi'\phi^3=0.
\end{align}

Throughout this article, the metric signature will be $\eta_{\mu\nu}=$diag$(-,+)$.
It is important to note that, if we consider the static solution in the two-dimensional space-time, $\phi=\phi(x_1)\equiv \phi(x)$, the contribution from the kinetic term to the equation of motion is $\partial_{\mu}(\frac{\partial L}{\partial(\partial_{\mu}\phi)})=(\frac{\partial L}{\partial\phi^{\prime}})^{\prime}=\mu^2(\frac{\partial|\phi^{\prime}|}{\partial\phi^{\prime}})^{\prime}$, which vanishes except of singular case $\phi^{\prime}=0$. Nevertheless, this singularity is actually removable, thus,  one can assign the zero value to its contribution to the equation of motion as well. Consequently, in two-dimensional case the pure cuscuton-like kinetic term yields only a trivial contribution to equations of motion independently of the form of the potential. Therefore, such a kinetic term does not contribute to the equation of motion. As a matter of fact, the Euler-Lagrange equation is completely described by the potential.

Indeed, analyzing the solution of (\ref{kink}), we get that 
\begin{align}
    \phi=M\,\exp{\bigg(\frac{25}{12}-\frac{1}{4\lambda}\bigg)},
\end{align}
that is, a constant.
Therefore, the model does not admit topological structures. 

\subsubsection{The kink-like solution}

To find a scenario where a cuscuton term develops a nontrivial contribution to the equation of motion, we define a model that includes both terms with canonical and non-canonical dynamics, namely, 
\begin{align}
	\label{cus_g}
    S=\int\, d^2x\, \bigg[\mu^2\sqrt{\partial_\mu\phi\partial^{\mu}\phi}+\frac{1}{2}\partial_\mu\phi\partial^\mu\phi-V(\phi)\bigg].
\end{align}

In this case, the equation of motion (\ref{kink}) describing the static field configurations, for our choice of the potential (\ref{part}) is
\begin{align}
    \label{kink1}
    \phi''-\frac{\lambda}{64\pi^2}\bigg\{\lambda\bigg[\text{ln}\bigg(\frac{\phi^2}{M^2}\bigg)-\frac{25}{6}\bigg]+\frac{1}{2}\bigg\}\phi'\phi^3=0.
\end{align}

Due to non-linearity of equation of motion, we must use a numerical method to investigate classical field solutions. Investigating topological field configurations, we will assume that
\begin{align}
    \phi(x\rightarrow -\infty)\to -1 \hspace{1cm}\text{and}\hspace{1cm} \phi(x\to\infty)\to 1.
\end{align}

\begin{figure}[ht!]
    \centering
    \includegraphics[scale=0.6]{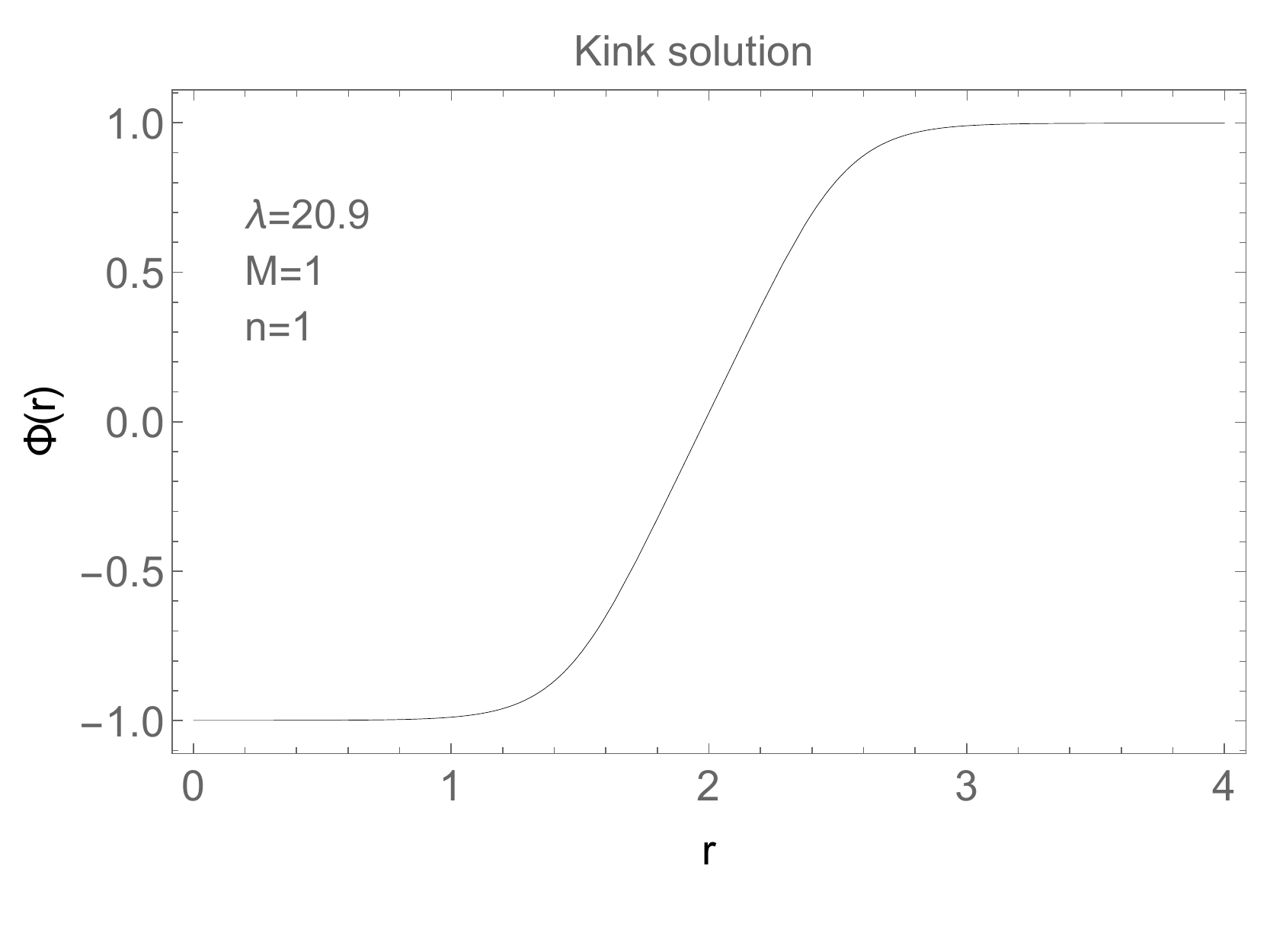}
    \vspace{-10pt}
    \caption{Kink-like solutions in the cuscuton model added to the term of the standard canonical dynamics in 2D.}
    \label{figK}
\end{figure}

Studying numerically the solution of Eq. (\ref{kink}), we obtain, by interpolation, the kink-like solution presented in Fig. \ref{figK}. 

In this case, the energy density, at $\mu=1$, is
\begin{align}
\label{E}
\mathcal{E}=\frac{1}{2}\phi'^2+\phi'-\frac{\lambda^2\phi^4}{256\pi^2}\bigg[\text{ln}\bigg(\frac{\phi^2}{M^2}\
\bigg)-\frac{25}{6}\bigg].
\end{align}
We note that the linear term $\phi^{\prime}$ here is a consequence of the cuscuton contribution. 

We refer to the structures as kink-like solutions due to the scalar field profile and energy density, i.e.  the structures that arise in the cuscuton-like theory have a similar profile as a kink structure coming from a $\phi^4$ theory.

\begin{figure}[ht!]
    \centering
    \includegraphics[scale=0.6]{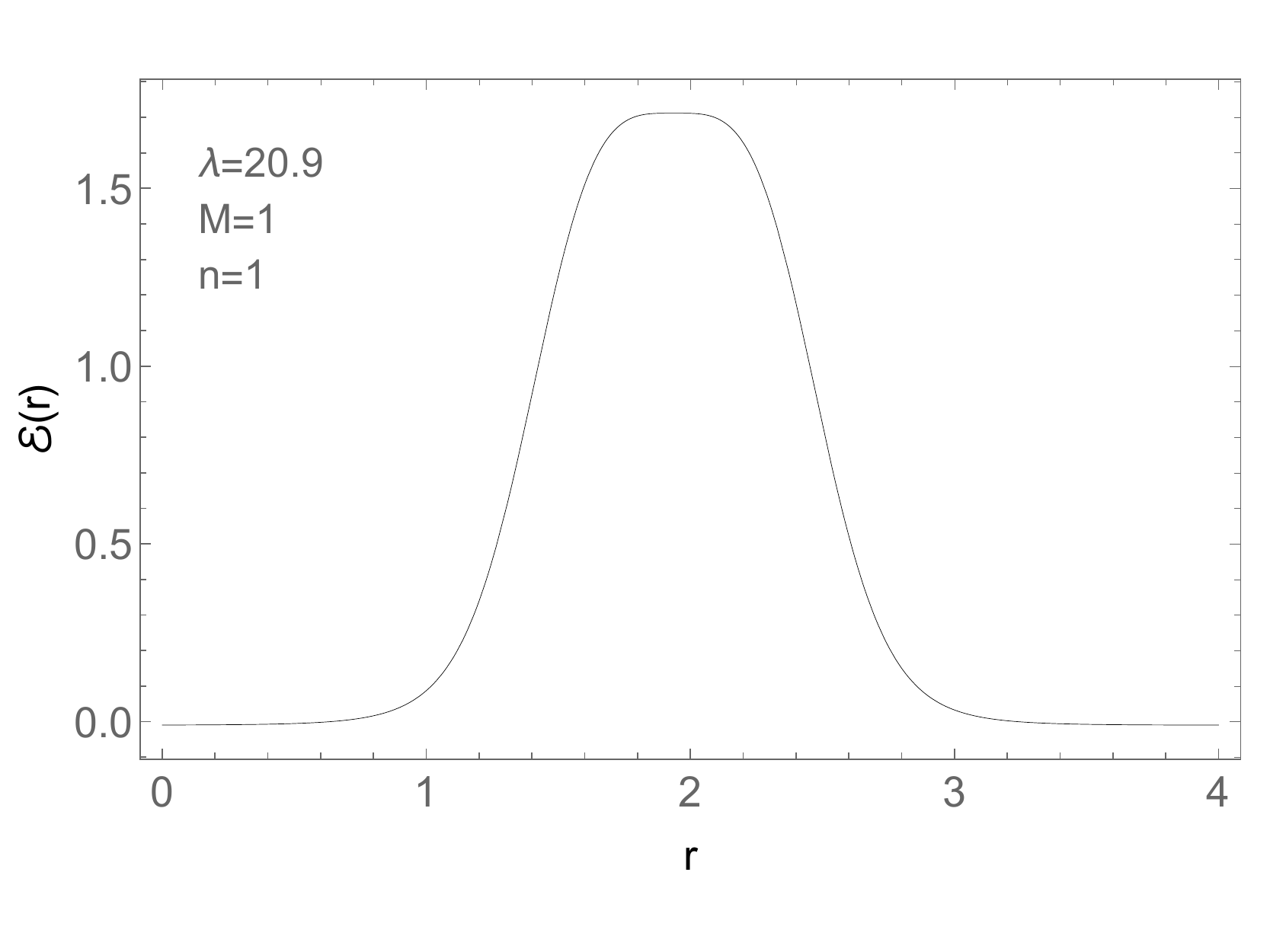}
    \vspace{-10pt}
    \caption{Energy density of the field $\phi$.}
    \label{figke}
\end{figure}

Let us analyze the behavior of the energy density of the kink-like solution in Fig. \ref{figK}. For this analysis, we considered energy (\ref{E}) and investigated the behavior of energy for the field of Fig. \ref{figK}. The result we obtained is shown in Fig. \ref{figke}. 

As expected, adding the canonical kinetic term to the $2D$ cuscuton model solves the problem and allows the emergence of topological structures kink-like. Here it is interesting to mention that by adjusting the parameters $M$, and $\lambda$, we will obtain other kink-like profiles centered in other regions of space. In our simulation, we show the kink-like structure is centered at $r=2$. By analysis, we see that the solutions of  Eq. (\ref{kink1}) are type $\phi\propto \tanh (r-r_0)$, where the value of $r_0$ is associated with the parameters $M$ and $\lambda$. As a consequence, we observed that the energy associated with this structure behaves like ${\rm sech}(r-2)^2$. This result is a direct consequence of the influence of the addition of the canonical dynamics in the cuscuton-like model. Some interesting discussions of this class of solutions can be found in Refs. \cite{Manton,Speight,Kodama}.

Within that study of kink-like structures, we concluded that the pure non-canonical cuscuton kinetic term yields only a trivial contribution independently of the form of the potential. Nontrivial impacts of the square root term probably can arise for generalized cuscuton terms like $f(\phi)\sqrt{\partial^{\mu}\phi\partial_{\mu}\phi}$ or for cuscuton-like models including an additive contribution of the canonical kinetic term. Some implications of such terms have been considered in Ref. \cite{BMP}, and their detailed study will be carried out elsewhere. It is natural to expect that other potentials would display similar behaviour.

 \begin{center}
\textit{{\bf Linear excitation of the kink-like structures}}
\end{center}

In the background of the kink-like structure, the dispersion states are found as solutions to the nontrivial equation of motion of kink-like fluctuations. If we denote the solution of the stationary matter field by $\phi_k(r)$, the fluctuation field $\psi_(r,t)$ is
\begin{align}
    \delta \phi(r,t)=\phi^{(k)}(r,t)-\delta\phi(r,t),
\end{align}
let us assume that the vacuum expected value is much smaller than the fluctuations in the field. To study fluctuations, it is convenient to consider $\psi(x,t)=h(x)\cos{\omega t}$. Considering the linear terms of the theory, we arrive at the equation
\begin{align}\label{Schr}
    -\frac{d^2 h(r)}{d r^2}+U(r)h(r)=\omega^2 h(r),
\end{align}
the equation (\ref{Schr}) is a Schr\"{o}dinger-like equation with
\begin{align}
    U(r)=\frac{d^2V^{(k)}}{d\phi^2}\bigg\vert_{\phi=\phi^{(k)}}.
\end{align}
being the stability potential referred to as the quantum-mechanical potential. For our model, the stability potential is
\begin{align}\label{pesta}
    U(r)=\frac{\lambda^2}{64\pi^2}\bigg[3\text{ln}\bigg(\frac{\phi^{(k)^2}}{M^2}-\frac{25}{6}\bigg)+\frac{7}{2}\bigg]\phi^{(k)^{2}}\phi^{'\, (k)^2}.
\end{align}

\begin{figure}[ht!]
    \centering
    \includegraphics[scale=0.6]{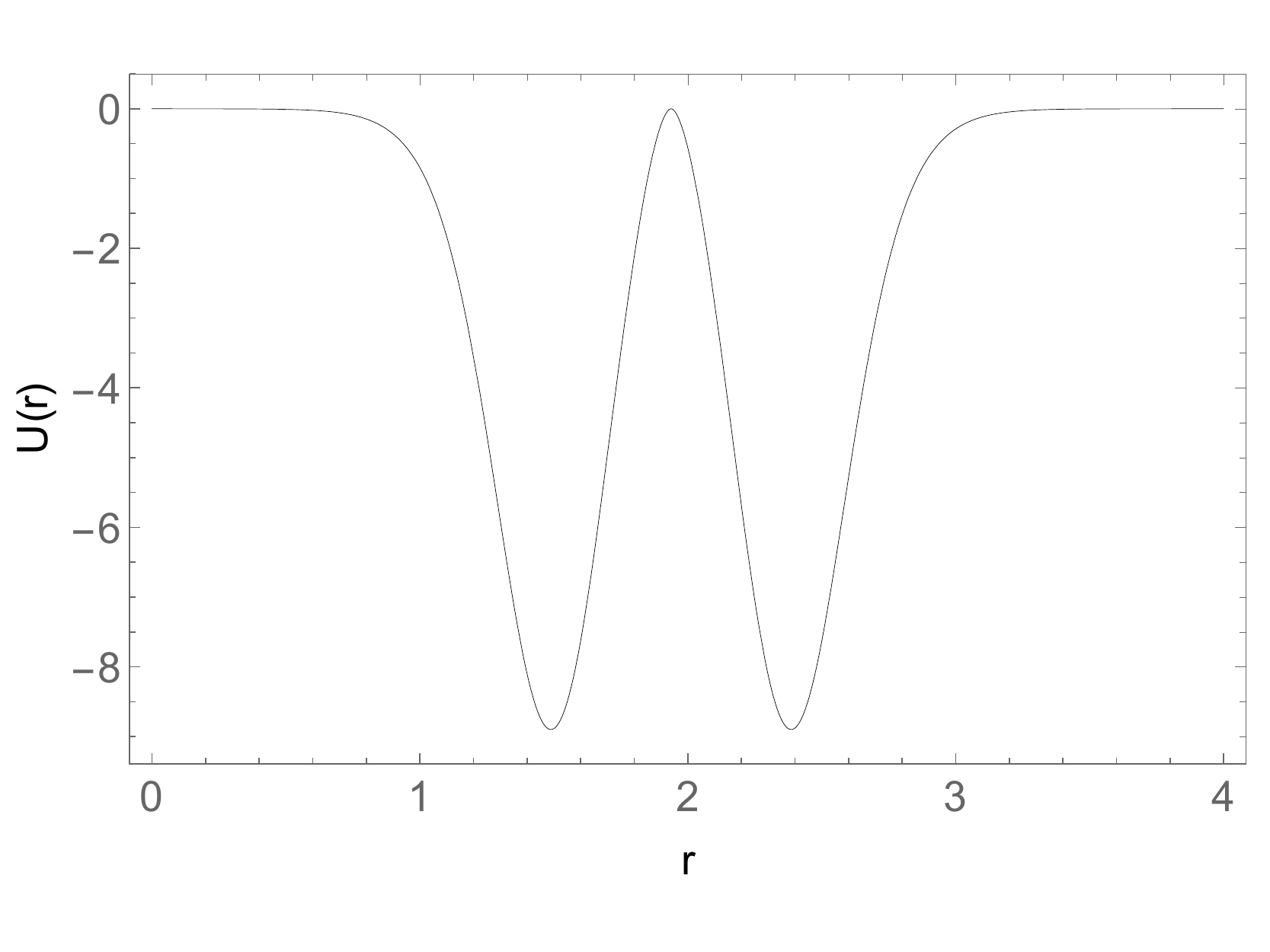}
    \vspace{-10pt}
    \caption{Profile of potential stability when $M=1$ and $\lambda=20.9$}
    \label{figpot}
\end{figure}

To investigate the excitation spectrum of our kink-like solutions, we analyze the stability potential (\ref{pesta}). The stability potential profile is shown in fig. \ref{figpot}.

\begin{figure}[ht!]
    \centering
    \includegraphics[scale=0.6]{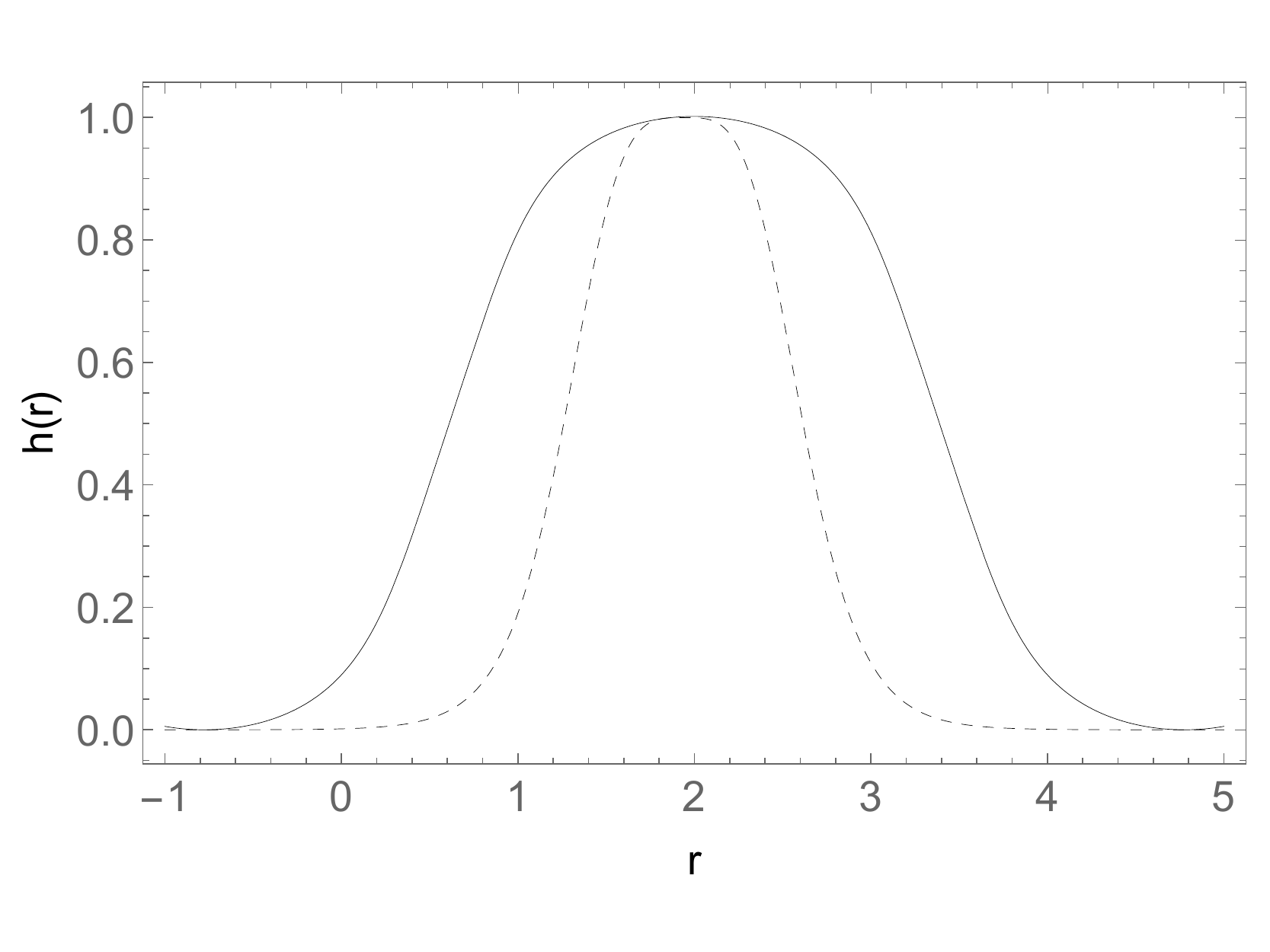}
    \vspace{-10pt}
    \caption{Profiles of the eigenfunction of zero-mode (translational mode, continuous line), and eigenfunction of the Eq. (\ref{Schr}) with eigenvalue $\omega_0=0.663$ (dashed line)}
    \label{figesp}
\end{figure}

Perceive that the stability potential is one symmetrical double-well centered at $r=2$, with $U(r)<0$. It is worth mentioning that the confining potential is located at the position of the structure. This behavior is reflected in the profile of the numerical solution of the Eq. (\ref{Schr}), and leads us to investigate the translation modes of the structure. Using the numerical interpolation method with steps of $10^{-5}$, we show the ground state of the excitation spectrum (see Fig. \ref{figesp}).

Note that there are no negative values of the eigenvalue of the excitation spectrum of our kink-like structures. Furthermore, using arguments from supersymmetric quantum theory, it is concluded that $\frac{d\phi^{(k)}}{dr}$ is the corresponding eigenfunction to the eigenvalue $\omega_0=0$ (the calculation of this proof is immediate by acting the annihilation operator on the ground state of the spectrum, for more details see Ref. \cite{Vachaspati}). This result implies that kink or kink-like structures must have a zero (translational) mode, see for example Ref. \cite{BelenGZ}.

To achieve our purpose, the numerical interpolation method was used to investigate the solution of Eq. (\ref{Schr}). By numerical interpolation, the only frequency of the spectrum arises when $\omega_{0}=0.663$. Its corresponding eigenfunction is shown in Fig. \ref{figesp}. Analyzing the vibrational modes, is it noted the absence of these eigenstates. This result is interesting, as it suggests that resonance phenomena probably do not occur in kink-antikink scattering in our structures. However, it is important to note that the phenomenon of resonance is possible even in the absence of vibrational modes, see Ref. \cite{BelenGZ}. We hope in future works to investigate the scattering of these structures to analyze the possibility of the existence of the resonance phenomenon.

\subsection{Vortex solutions gauged by the Maxwell field}

\subsubsection{Description of the theory}

Let us also motivate ourselves now by the theory presented above, to investigate the vortex structures associated with cuscuton-like \cite{cuscos} dynamics. To investigate the vortex structures it is necessary to couple a gauge field for the investigation of these field configurations. In this way, we will consider the model defined by the action,
\begin{align}
\label{CuscutonM}
    S=\int\, d^3x \bigg[\mu^2\sqrt{ \,\overline{D_\mu\phi}\cdot D^{\mu}\phi}-\frac{1}{4}F_{\mu\nu}F^{\mu\nu}-V(\vert\phi\vert)\bigg],
\end{align}
where again, the potential is chosen in the form (\ref{pot}).

Let us make our definitions.  The field $\phi$ is the complex scalar field.  Here, an overline denotes the complex conjugation of the field.  The Abelian gauge field is denoted by $A_{\mu}$.  The electromagnetic tensor is described by $F_{\mu\nu}$, which is defined as
\begin{align}
    F_{\mu\nu}=\partial_\mu A_\nu-\partial_\nu A_\mu.
\end{align}

Also, the  covariant  derivative is defined as $D_\mu=\partial_\mu+ieA_\mu$.   The  metric signature is $\eta_{\mu\nu}=$diag$(-,+,+)$ and the natural system of units has $\hbar=c=1$.

Investigating the equation of motion of the field configurations, we obtain that 
\begin{align}
    \label{motion}
     D_{\mu}\bigg(\frac{D^{\mu}\phi}{\sqrt{\vert D_\nu\phi D^\nu\phi\vert}}\bigg)+\frac{\phi}{\vert\phi\vert}V_{\vert\phi\vert}=0,
\end{align}
and
\begin{align}
    \partial_\mu F^{\mu\nu}=J^\nu.
\end{align}

For this non-canonical field dynamics, the current $J^\nu$ takes the form 
\begin{align}
    J_\nu=\frac{ie}{\sqrt{2\vert\partial_\mu\phi\partial^\mu\phi\vert}}(\overline{\phi}D_{\nu}\phi-\phi\overline{D_{\nu}\phi}).
\end{align}

Let us consider the translational symmetry of space-time to build the energy-momentum tensor, namely,
\begin{align}
    T_{\mu\nu}=\frac{1}{2}\sqrt{\overline{D_\mu\phi}D_\nu\phi}+\frac{1}{2}\sqrt{D_\mu\phi \overline{D_\nu\phi}}+F_{\mu\lambda}F^{\lambda}\,_{\nu}-\eta_{\mu\nu}\mathcal{L}
\end{align}

This allows us also to particularize our solutions to the case of the rotational symmetry. Therefore,  an ansatz can be 
\begin{align}
\label{AnC}
    \phi=f(r)\text{e}^{in\theta} \hspace{0.5cm} \text{and} \hspace{0.5cm} \textbf{A}=-\frac{1}{er}[a(r)-n]\hat{e}_{\theta},
\end{align}
where to investigate topological field conditions, we must assume that 
\begin{align}
\label{Cond}
    &f(0)=0, \hspace{2cm} f(\infty)=\nu,\\ \label{Cond1}
    &a(0)=n, \hspace{2cm} a(\infty)=0.
\end{align}
Here it is important to mention that $n\in\mathbb{Z}$ and it is also known as the winding number of the model. Meanwhile, $\nu$ is the vacuum expected value (v.e.v.). 

For this particular ansatz, we observe that the vortex structures are purely magnetic, with the field given by 
\begin{align}
\label{Bfield}
    \textbf{B}=-F_{12}=\nabla\times\textbf{A}=-\frac{a'(r)}{er}.
\end{align}

It is important to mention that these topological structures have a magnetic flux that is given by 
\begin{align}
    \Phi_B=\int_S\int\, \textbf{B}\cdot d\textbf{S}=-\int_{0}^{2\pi}\int_{0}^{\infty}F^{12}rdrd\theta=-\frac{2\pi}{e}[a(\infty)-a(0)].
\end{align}

Therefore, 
\begin{align}
    \label{FluxCusc}
    \Phi_B=\frac{2\pi n}{e}.
\end{align}

In this way, we are investigating quantized magnetic topological structures, i. e., structures that exhibit a quantized magnetic flux. 

Considering the behavior of the fields (\ref{AnC}), we have the energy density of the model written as 
\begin{align}
    \mathcal{E}=\sqrt{f'(r)^2+\frac{f(r)^2a(r)^2}{r^2}}+\frac{a'(r)^2}{e^2r^2}+\frac{\lambda^2f(r)^4}{256\pi^2}\bigg[\text{ln}\bigg(\frac{f(r)^2}{M^2}\bigg)-\frac{25}{6}\bigg].
\end{align}

To find finite energy configurations in the model, the energy functional can be rearranged as 
\begin{align} \nonumber
    \mathcal{E}=&\sqrt{\bigg(f'(r)-\frac{f(r)a(r)}{r}\bigg)^2+\frac{2f'(r)f(r)a(r)}{r}}+\bigg\{\frac{a'(r)}{er}\mp \frac{\lambda f(r)^2}{8\pi}\bigg[\text{ln}\bigg(\frac{f(r)^2}{M^2}\bigg)-\frac{25}{6}\bigg]^{1/2}\bigg\}\\ 
    &\pm \frac{1}{4e\pi}\frac{\lambda a'(r)f(r)^2}{r}\bigg[\text{ln}\bigg(\frac{f(r)^2}{M^2}\bigg)-\frac{25}{6}\bigg]^{1/2}.
\end{align}

In order to study the localized structures of the model, we will assume that the fields obey the following equalities, 
\begin{align}
    \label{e1}
    f'(r)=\frac{f(r)a(r)}{r}, 
\end{align}
and
\begin{align}
\label{e2}
    \frac{a'(r)}{er}=\pm \frac{\lambda f(r)^2}{8\pi}\bigg[\text{ln}\bigg(\frac{f(r)^2}{M^2}\bigg)-\frac{25}{6}\bigg]^{1/2}.
\end{align}

Looking at (\ref{e1}) and (\ref{e2}), we can clearly see that at the limit of $\lambda\rightarrow 0$, the gauge field assumed a constant behavior, i. e., $a(r)=a_0$. Meanwhile, we have $f(r)\propto r^{a_0}$, which gives us non-topological field configurations.

In fact, for all field configurations that respect Eqs. (\ref{e1}) and (\ref{e2}), the energy density is reduced to 
\begin{align}
\label{E_BP}
    \mathcal{E}_{BPS}=&\sqrt{\frac{2f'(r)f(r)a(r)}{r}}\pm \frac{1}{4e\pi}\frac{\lambda a'(r)f(r)^2}{r}\bigg[\text{ln}\bigg(\frac{f(r)^2}{M^2}\bigg)-\frac{25}{6}\bigg]^{1/2}.
\end{align}

By convention, $\mathcal{E}_{BPS}$ can be treated as the model's BPS energy \cite{B,PS,L}.

\subsubsection{Numerical results}

\begin{figure}[ht!]
    \centering
    \includegraphics[scale=0.45]{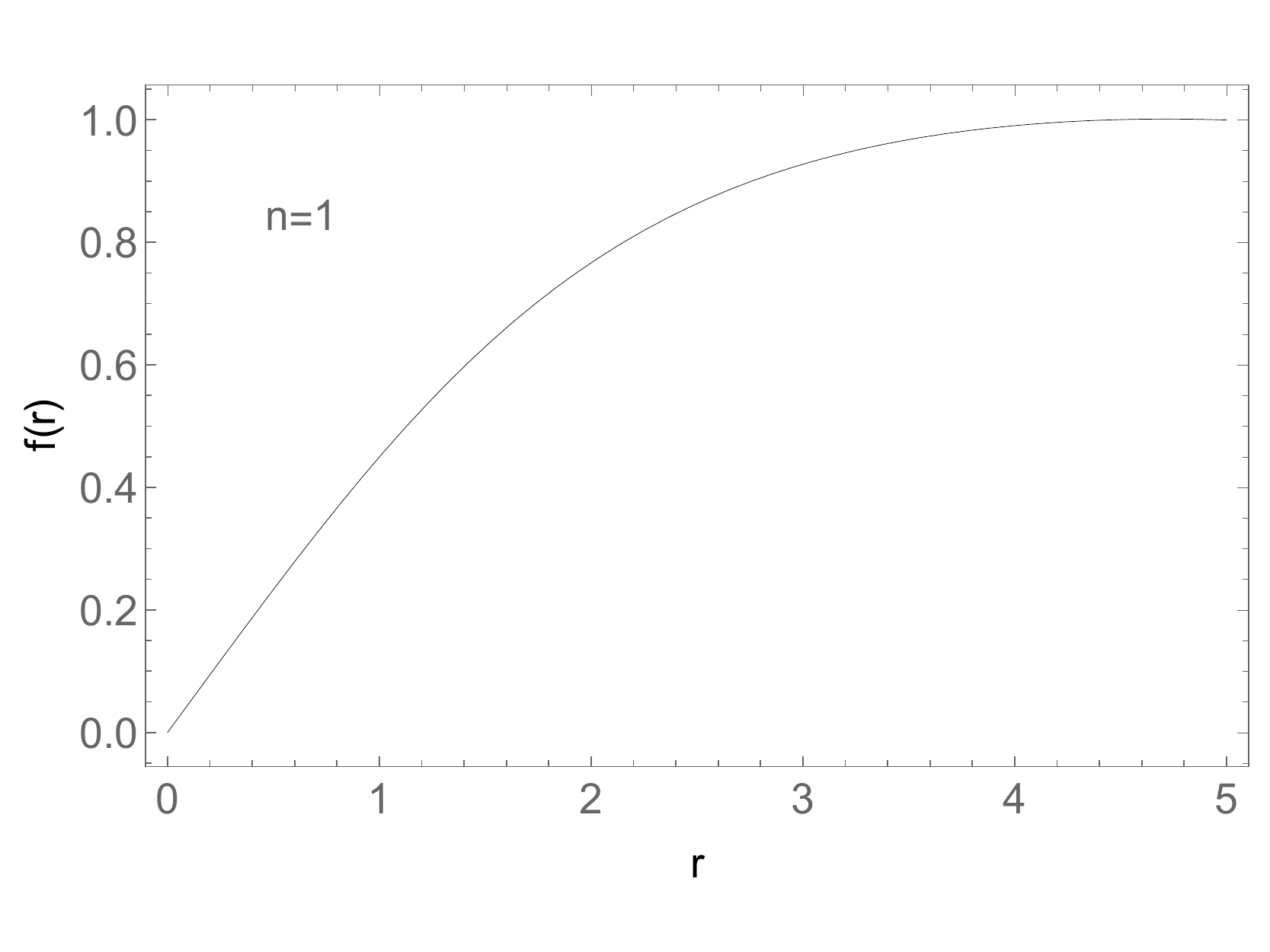}
    \includegraphics[scale=0.45]{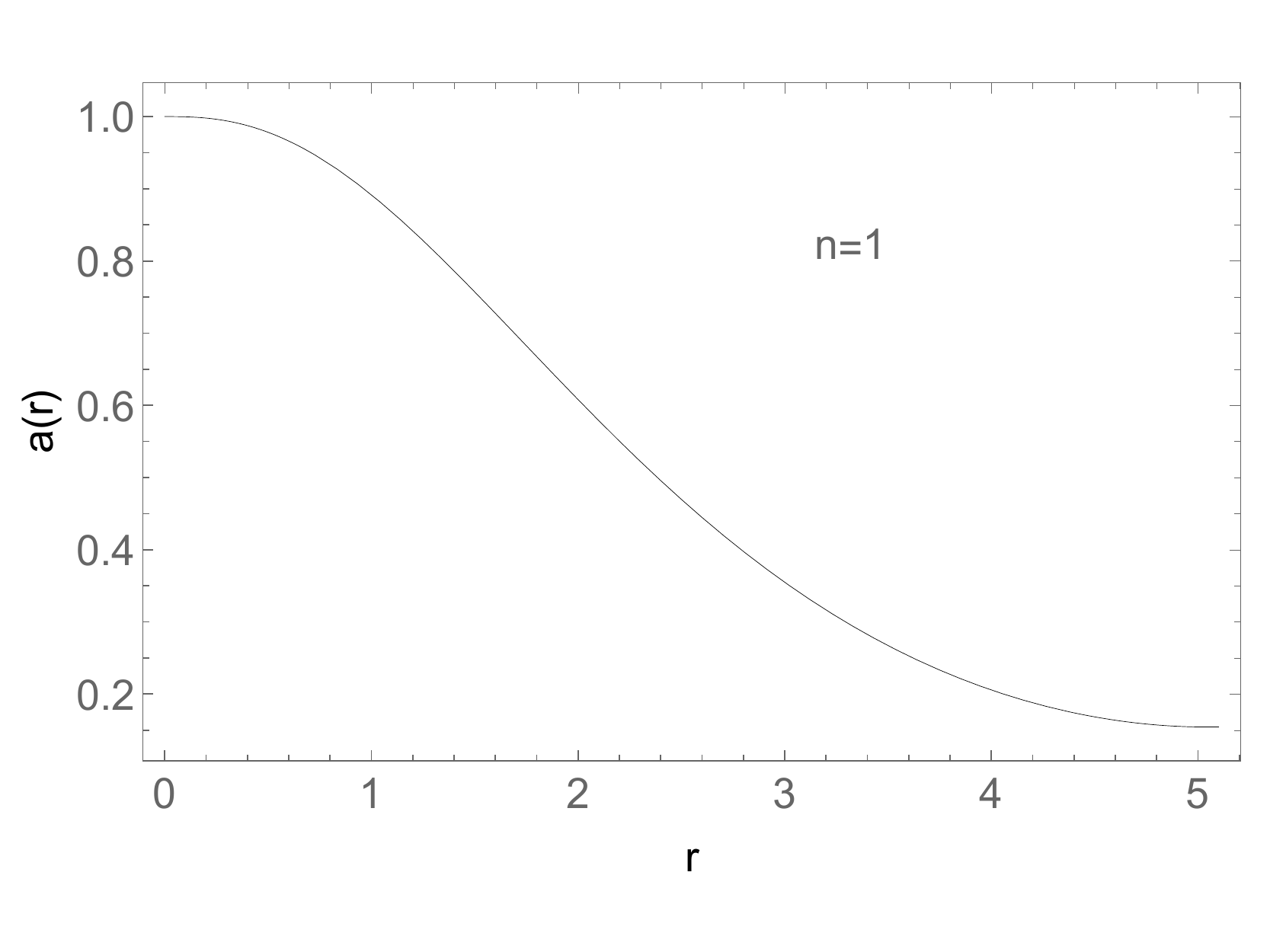}\\
    (a) \, \, \, \, \, \, \, \hspace{5cm} \, \, \, \, \, \, \, (b)
    \caption{(a) Behavior of the variable field $f(r)$. (b) Solution of the gauge field associated with field $f(r)$. In both solutions we assume that $e=\lambda=M=1$.}
    \label{fig1}
\end{figure}

In fact, the chosen case significantly simplifies the configurations of fields that have energy located in the cuscuton-like model. However, to investigate the solutions of Eqs. (\ref{e1}) and (\ref{e2}), a numerical method is required. For our case, we use a numerical interpolation to study the field configurations expressed by Eqs. (\ref{e1}) and (\ref{e2}). The numerical solutions for the field variables $f(r)$ and $a(r)$ are illustrated in Figs. \ref{fig1}(a) and \ref{fig1}(b), respectively.

\begin{figure}[ht!]
    \centering
    \includegraphics[scale=0.6]{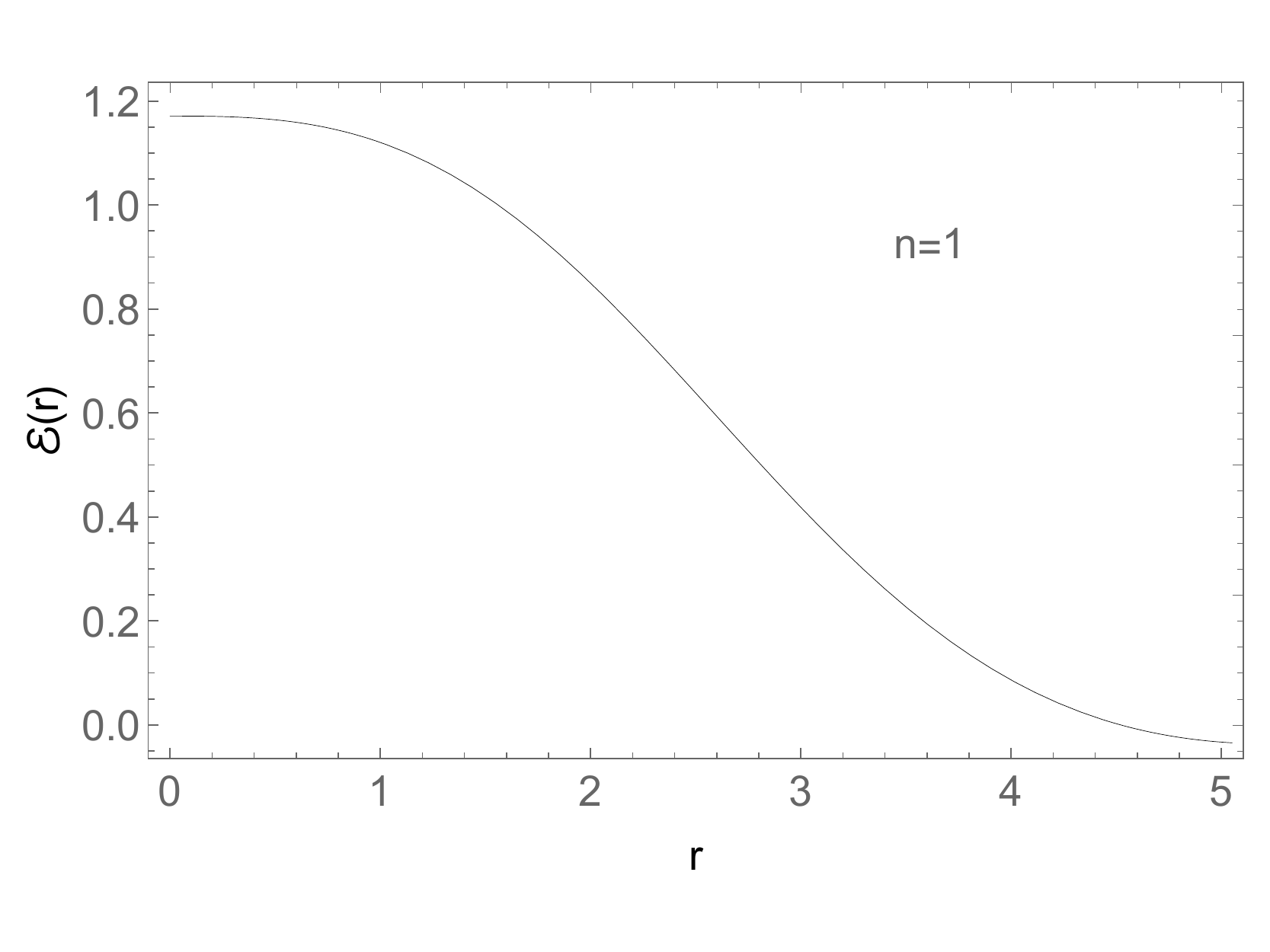}
    \caption{Energy density associated with complex scalar and gauge field solutions when $e=\lambda=M=1$.}
    \label{fig2}
\end{figure}

The vortex solutions with the field configurations shown in Fig. \ref{fig1} have localized energy. This is clearly seen when we investigate the energy density (\ref{E_BP}) for the behavior of $f(r)$ and $a(r)$ in Fig. \ref{fig1}. The result for the energy density of the model is shown in Fig. \ref{fig2}. 

\begin{figure}[ht!]
    \centering
    \includegraphics[scale=0.6]{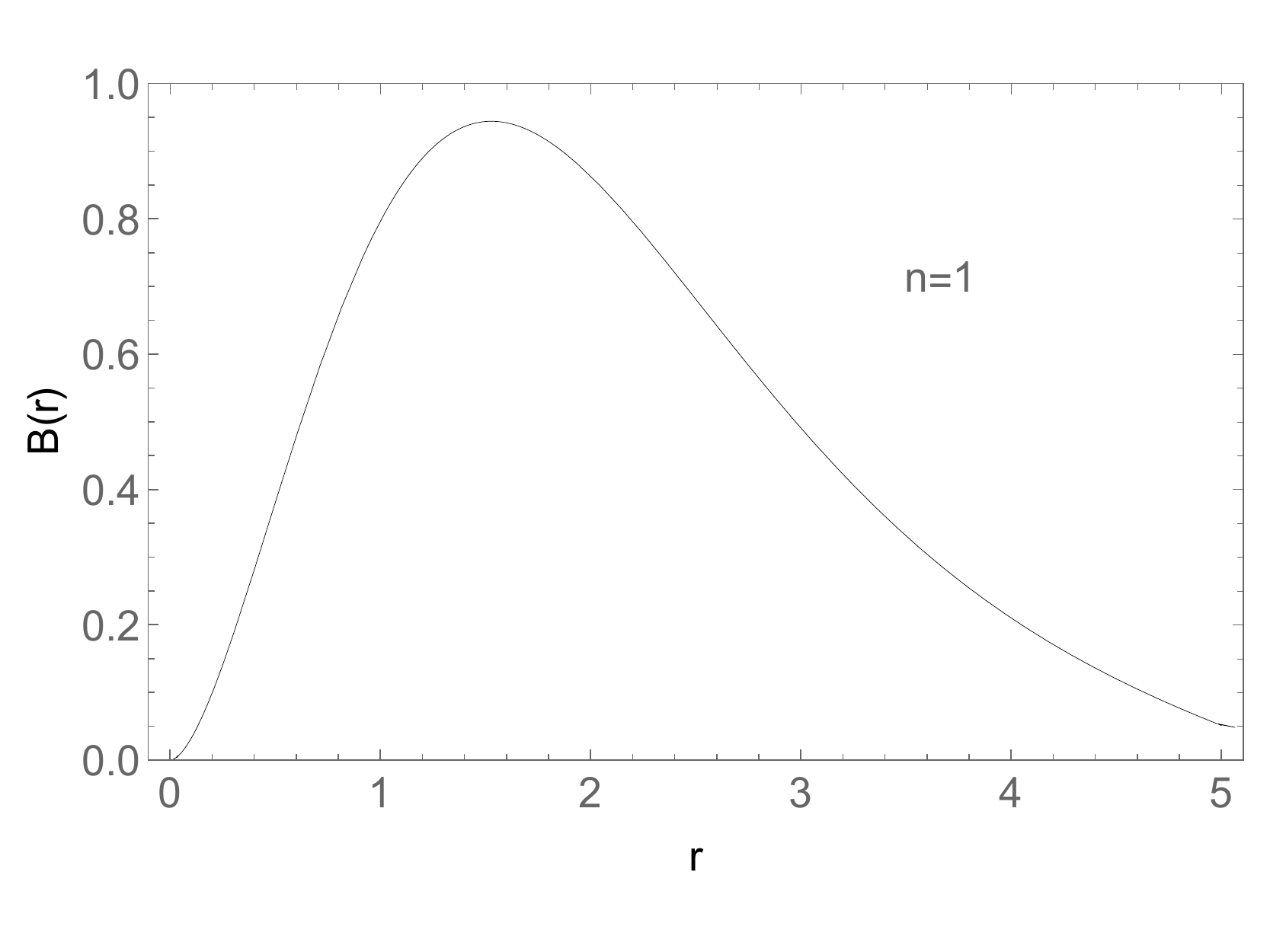}
    \caption{Magnetic field associated the vortex solution when $e=\lambda=M=1$.}
    \label{fig3}
\end{figure}

The magnetic field that generates the flux of the vortex is found considering the solutions in Fig. \ref{fig1} and the Eq. (\ref{Bfield}). By interpolation, the behavior of the magnetic field is demonstrated in Fig. \ref{fig3}.

It is easy to check that the considerations carried out throughout this section, continue to be valid, if we consider a set of $\phi_i$ fields, with $i=1..N$, so that only one of these fields, say $\phi_1$, differs from zero, while all other ones are equal to zero. In this case the solutions obtained in this section continue to be valid for the $\phi_1$ field. This is the simplest solution for the cuscuton theory generalized to the case of $N$ scalar fields which is very interesting from the perturbative viewpoint. Indeed, as we will argue in the next section, the corresponding extension of the cuscuton will admit a consistent quantum treatment based on using of the $1/N$ expansion. At the same time, it is natural to expect that other exact solutions consistent in the presence of many scalar fields $\phi_i$ can be obtained as well. Before starting the next section, where we will discuss quantum perturbative spects of the cuscuton,  let us emphasize that the low-dimensional single field solutions we obtained above represent themselves as the first steps for inspecting the corresponding quantum theory which we begin to consider now.

\section{Comments on perturbative dynamics of the cuscuton}

While, as it was demonstrated in the introduction, the main attention to the cuscuton model is motivated to its classical properties, i. e. exact solutions, quantum corrections in this theory call a certain interest motivated, clearly, by the highly unusual form of the kinetic term establishing the question about a possibility of consistent application of the perturbative formalism of the quantum field theory. It is clear that, to solve this question, one must define a field theory model equivalent in a certain limit to a cuscuton theory,  but written in the standard form, i.e. its Lagrangian looks like a sum of the conventional kinetic term $\frac{1}{2}\phi_i\hat{T}_{ij}\phi_j$, with $\hat{T}_{ij}$ some differential operator acting on dynamical fields $\phi_j$, and an interaction term presented as a power series in fields and derivatives, so that in some limit (i.e. when some constant parameters tend to zero or to the infinity) or after integrating out some fields, one could arrive at the cuscuton model. For the theory with the Lagrangian ${\cal L}=\frac{1}{2}\phi_i\hat{T}_{ij}\phi_j+\ldots$, one evidently can apply the standard pertubative methodology.


First, one can assume the naive manner to generalize the cuscuton model in the Born-Infeld style, proposing the action
\bea
\label{cus1}
S_1=\int d^4x \bigg[\mu^4\sqrt{1+\frac{\partial_{\mu}\phi\partial^{\mu}\phi}{\mu^4}}-V(\phi)\bigg],
\eea
where, after all calculations, we should consider the small $\mu$ limit in order to have the first, field independent term to be suppressed. We note that this Born-Infeld-like action is related with the cuscuton just in the same manner as the Born-Infeld electrodynamics is related to the Nielsen-Olesen gauge theory, discussed in Ref. \cite{NieOl} (see also references therein). However, this model cannot be treated as a perturbatively consistent generalization of the cuscuton model. Indeed, the cuscuton action corresponds to small (not large!) $\mu^2$ limit of Eq. (\ref{cus1}), for which the unit under the square root is suppressed in comparison with $\frac{\partial_{\mu}\phi\partial^{\mu}\phi}{\mu^4}$, to ensure arising of the classical cuscuton action (\ref{cus}), and, by dimensional reasons, as higher the order in fields and derivatives, as higher will be the negative degree of small $\mu$. Therefore, the series in $\mu$ diverges, hence the presentation of the cuscuton action in the form (\ref{cus1}) is perturbatively inconsistent.

The more appropriate way for inserting the cuscuton model in a perturbative context is based on the sigma-model-style approach. We start with the action
\bea
\label{cus2}
S_2=\int d^4x\Big(\frac{1}{2}(\Sigma \partial^{\mu}\phi\partial_{\mu}\phi+\frac{\mu^4}{\Sigma})-V(\phi)
\Big),
\eea
which, after eliminating $\Sigma$ through its classical equations of motion, reproduces the initial cuscuton action (\ref{cus}). Therefore, our theory is dynamically equivalent to the cuscuton model. Further, to  avoid negative degrees of any fields, we can replace $\Sigma\to e^{\sigma}$, since $\Sigma$ is dimensionless:
\bea
\label{cus2a}
S'_2=\int d^4x\Big(\frac{1}{2}(e^{\sigma} \partial^{\mu}\phi\partial_{\mu}\phi+\mu^4e^{-\sigma})-V(\phi)
\Big).
\eea
Now, we perform the sigma-model-like extension of this theory, by assuming that it describes, instead of one field $\phi$, $N$ scalar fields $\phi_i$, $i=1...N$, which allows to employ  $1/N$ expansion as it is done for the nonlinear sigma model \cite{GN}, so, we promote the theory (\ref{cus2a}) to
\bea
\label{cus2b}
S'_2=\int d^4x\Big(\frac{1}{2}(e^{\sigma} \sum\limits_{i=1}^N\partial^{\mu}\phi_i\partial_{\mu}\phi_i+\mu^4e^{-\sigma})-V(\phi)
\Big).
\eea
Eliminating the $\sigma$ field with its equation of motion, we arrive at the many-field cuscuton kinetic term $\sqrt{\sum\limits_{i=1}^N\partial^{\mu}\phi_i\partial_{\mu}\phi_i}$. Therefore, the theory we consider is effectively the many-field extension of the cuscuton model, with the multiplicity of fields is necessary to apply the machinery of the $1/N$ expansion.

The propagator of $\phi_i$ is usual:
\bea
\braket{\phi_i\phi_j}=\frac{\delta_{ij}}{k^2-m^2}.
\eea

We choose the usual self-coupling potential
\bea
V(\phi)=\frac{m^2}{2}\sum\limits_{i=1}^N\phi_i\phi_i+\frac{\lambda}{4!}(\sum\limits_{i=1}^N\phi_i\phi_i)^2.
\eea

There will be two contributions to the two-point function of $\sigma$ field, that one involving three-point vertices and that one involving a four-point vertex.

\begin{figure}[ht!]
	\begin{centering}
		\includegraphics[height=2.5cm]{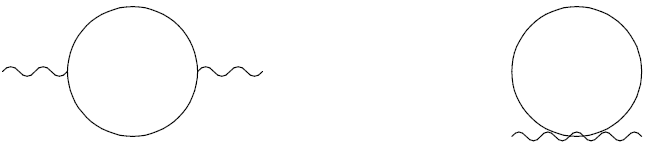}\\
		\hspace{1cm} a \hspace{6.5cm} b
		\par\end{centering}
	\caption{One loop contributions to the two-point functions of $\sigma$ field.}
	\label{ffey}
\end{figure}

These vertices, respectively, look like
\bea
V_3&=&\frac{1}{2}\sigma\sum\limits_{i=1}^N\partial^{\mu}\phi_i\partial_{\mu}\phi_i;\nonumber\\
V_4&=&\frac{1}{4}\sigma^2\sum\limits_{i=1}^N\partial^{\mu}\phi_i\partial_{\mu}\phi_i.
\eea

For the sake of simplicity, we suggest $V(\phi)=\frac{1}{2}m^2\sum\limits_{i=1}^N\phi_i\phi_i$, so, the $\phi_i$ are coupled only to the auxiliary field $\sigma$.
Hence the graphs $a, b$ depicted at Fig. \ref{ffey} yield following contributions to the effective Lagrangian:
\bea
\Gamma_a(p)=\frac{N}{8}\sigma(-p)\sigma(p)\int\frac{d^4k}{(2\pi)^4}\frac{k^2(k+p)^2}{(k^2-m^2)[(k+p)^2-m^2]};
\nonumber\\
\Gamma_b(p)=\frac{N}{4}\sigma(-p)\sigma(p)\int\frac{d^4k}{(2\pi)^4}\frac{k^2}{k^2-m^2}.
\eea

It is straightforward to show that although the exact result for the first integral is rather complicated due to need to integrate over Feynman parameters, for studying of renormalizability it is sufficient to use  the asymptotic form of the two-point function given by the sum $\Gamma_a+\Gamma_b$, just as in \cite{GN}:
\bea
\label{gsigma}
\Gamma_2&=&\Gamma_a+\Gamma_b\sim N\frac{\alpha}{2}\sigma(-p)(p^4+\beta m^4)\sigma(p),
\eea
where $\alpha,\beta$ are some numbers. So, we generated the higher-derivative contribution to the effective action. We should note nevertheless that it is divergent, so, the effective propagator of $\sigma$ is an inverse to this two-point function, i.e. $\braket{\sigma(p)\sigma(-p)}\propto\frac{1}{N}\frac{1}{p^4+m^4}$, but it is well defined only at $d=4-2\epsilon$ with $\epsilon\neq 0$. The superficial degree of divergence is 
\bea
\omega=4-E_{\phi}-N_d,
\eea
where $E_{\phi}$ is a number of external $\phi$  fields (it is always even), and $N_d$ is a number of derivatives applied to external fields, while the number of external $\sigma$ fields is arbitrary. One could think that renormalization of the theory is problematic. However, it follows from the structure of the classical action that the external $\sigma$ in any Feynman diagram will arise only within the exponential $e^{-\sigma}$. Therefore, all divergences can be of the form $e^{-n\sigma}$, $e^{-n\sigma}\sum\limits_{i=1}^N\partial^{\mu}\phi_i\partial_{\mu}\phi_i$, $e^{-n\sigma}(\sum\limits_{i=1}^N\phi_i\phi_i)^2$, with $n$ is non-negative integer (indeed, each vertex can carry only an integer number of external $e^{-\sigma}$). Alternatively, we can make resummation over background constant $\sigma$ and use the background-dependent propagator $\braket{\phi_i\phi_j}=\frac{1}{k^2e^{-\sigma}-m^2}\delta_{ij}$ which, for the constant background $\sigma$, allows to take care only on external $\phi_i$ fields while the $\sigma$ fields are already summed out. Moreover, if derivatives act on some external $\sigma$, it is easy to see that other external $\sigma$ fields in each vertex are summed to exponential. Even if we consider non-constant background $\sigma$, we can have only additional divergences $e^{-n\sigma}\partial^{\mu}\sigma\partial_{\mu}\sigma$, $e^{-n\sigma}(\partial^{\mu}\sigma\partial_{\mu}\sigma)^2$, and $e^{-n\sigma}\partial^{\mu}\sigma\partial_{\mu}\sigma\phi_i\phi_i$, with no other divergences.  Moreover, if the self-coupling of the $\phi_i$ is absent, $\lambda=0$, there will be no derivative-free vertices involving $\phi$ field. Therefore, we will have only two types of divergences: $e^{-n\sigma}$ and $e^{-n\sigma}\sum\limits_{i=1}^N\partial^{\mu}\phi_i\partial_{\mu}\phi_i$.

It is easy to find the effective potential $V^{(1)}$ of the field $\sigma$ treated as a purely external one, in the simplest case $\lambda=0$: in this case  the $V^{(1)}$ is given by
\bea
V^{(1)}=\frac{i}{2}N{\rm Tr}\ln(e^{\sigma}\Box+m^2).
\eea

We note that this contribution to the effective potential is a dominant term within the framework of $1/N$ expansion since it involves a complete loop of propagators of $\phi_i$ fields, and each such a loop yields $N$ factor while each propagator of the $\sigma$ field yields a $1/N$ factor. Doing the integral at $d=4-2\epsilon$, we find
\bea
V^{(1)}=-\frac{N}{64\pi^2}m^4e^{-2\sigma}\left(\frac{2}{\epsilon}+\gamma-1-2\ln\left(\frac{m^2e^{-\sigma}}{\mu^{\prime 2}}\right)\right)+O(\epsilon).
\eea
Here $\mu^{\prime}$ is a normalization parameter.
It is interesting to note that, first, the divergent part of the one-loop contribution to the effective potential does not match the form of the classical action, second, besides of the $e^{-2\sigma}$, we have as well a new (finite) term $\sigma e^{-2\sigma}$ due to the logarithmic contribution. In terms of the initial $\Sigma=e^{\sigma}$, we have
\bea
V^{(1)}=-\frac{N}{64\pi^2}\frac{m^4}{\Sigma^2}\left(\frac{2}{\epsilon}+\gamma-1-2\ln\left(\frac{m^2}{\mu^{\prime 2}}\right)+2\ln\Sigma\right)+O(\epsilon).
\eea
We note that there will be no other terms of the first order in $N$ in the effective action since this is the only contribution involving the propagators of $\phi_i$ only. To cancel the divergence, we must introduce the term $\frac{m^4}{\Sigma^2}$ to the initial action (\ref{cus2}).  The contributions to the effective action generated by propagators of $\Sigma$ will be suppressed by various degrees of the $\frac{1}{N}$ factor, entering the propagator of the $\sigma=\ln \Sigma$ corresponding to the quadratic action (\ref{gsigma}). Therefore, this modification of the action of $\Sigma$ will modify quantum corrections only in higher orders of $\frac{1}{N}$ expansion.
	
It is interesting to note that, if we assume this term to be added with a multiplier $\varepsilon\ll 1$, in order to treat this term as a small one, the modified sigma-model-like cuscuton action becomes
\bea
\label{cus2d}
S_2=\int d^4x\Big(\frac{1}{2}(\Sigma \sum\limits_{i=1}^N\partial^{\mu}\phi_i\partial_{\mu}\phi_i+\frac{\mu^4}{\Sigma})+\frac{\varepsilon m^4}{\Sigma^2}-V(\phi)
\Big).
\eea

Now, let us obtain equations of motion for $\Sigma$ corresponding to this action. We have
\bea
\label{eqmodmot}
\sum\limits_{i=1}^N\partial^{\mu}\phi_i\partial_{\mu}\phi_i-\frac{\mu^4}{\Sigma^2}-2\frac{\varepsilon m^4}{\Sigma^3}=0.
\eea

In principle, the $\Sigma$ can be found from this equation through the Cardano formula for roots of a cubic equation. However, it is better to obtain $\Sigma$ as a series in $\varepsilon$:
\bea
\label{series}
\Sigma=\frac{\mu^2}{|\partial\phi|}+\varepsilon\Sigma_1+O(\varepsilon^2),
\eea
where $|\partial\phi|\equiv\sqrt{\sum\limits_{i=1}^N\partial^{\mu}\phi_i\partial_{\mu}\phi_i}$. We know that at $\varepsilon=0$, one has $\Sigma=\frac{\mu^2}{|\partial\phi|}$, and substituting this result to the equations of motion in $\varepsilon=0$ case, one recovers the initial cuscuton action (\ref{cus}).  The $\Sigma_1$ is a first-order approximation to be found: implementing the expansion (\ref{series}) into the equation of motion (\ref{eqmodmot}), one obtains, in the first order in $\varepsilon$, that $\Sigma_1=\frac{m^4}{\mu^4}$. Thus, we substituite $\Sigma=\frac{\mu^2}{|\partial\phi|}+\varepsilon\frac{m^4}{\mu^4}$ to the action (\ref{cus2a}) and find the action depending only on $\phi_i$ and their derivatives:
\bea
\label{sfin}
S_2[\phi]=\int d^4x \Big[\frac{1}{2}[(m^2+\frac{\mu^4}{m^2})\sqrt{\sum\limits_{i=1}^N\partial^{\mu}\phi_i\partial_{\mu}\phi_i}+\frac{\varepsilon m^4}{\mu^4}{\sum\limits_{i=1}^N\partial^{\mu}\phi_i\partial_{\mu}\phi_i}]-V(\phi)\Big],
\eea
i.e., the modified theory (\ref{cus2d}) yields a sum of the cuscuton term with a modified constant multiplier and the usual kinetic terms, plus the potential. This result is a reminiscence of the discussion in the first part of section II, where we argued that the usual additive kinetic term is necessary to provide a nontrivial impact of the cuscuton term in equations of motion. We note that the usual kinetic term here is suppressed due to the $\varepsilon$ factor, hence we conclude that our theory (\ref{cus2d}) indeed reproduces a sum of the cuscuton action and small terms.

\section{Conclusion}

In this work, we constructed cuscuton models in $2D$ and $3D$ to study the existence of topological structures with the interaction term derived of the pertubative theory in the cuscuton model. Initially, the existence of kink solutions in the $2D$ model was investigated, and the nontrivial solutions were shown to arise if we have both non-canonical and canonical kinetic terms in the action. Then, we studied Maxwell's vortices subject to the nonpolynomial potential. In our model, the gauge field is described by Maxwell's term. Starting our investigation, we observed that the cuscuton-like model only admits non-topological solutions. However, when adding a canonical kinetic term to the model, we notice that the model starts to admit the arise topological structures. In fact, the existence of kink structures in the $2D$ model was expected since the corrected nonlinear potential assumes a behavior similar to a $\phi^4$ theory. We observed, by numerical analysis, that the scalar field (in case more general) assumes a behavior like $\phi\propto\tanh(x)$. Thus, an immediate consequence is that the energy density of the model starts to assume a profile similar to a Gaussian curve, namely $\mathcal{E}\propto$ sech$(x)^2$.

For an analysis of the vortex structure, we note that the analyzed vortices have a critical gauge field around the origin, i. e., when $r\to 0$. On the other hand, it is notable that the gauge field of these structures generates a null magnetic field in the near of $r=0$. However, it quickly reached a maximum intensity value of $r_c$ ($r_c$ is the value of $r$ that the scalar field reach the v.e.v.). It is important to mention that the vortex structures that admit this behavior are said to be ringlike vortex \cite{L}. Furthermore, all the vortex structures investigated so far have a quantized magnetic flux given by $2\pi n/e$.

In fact, due to the behavior of the field variable $f(r)$, the energy settings of the vortex are localized and have the profile of a function ${\rm sech}(r)^2$. It is worth noting that despite the localized characteristic of the structures in both cases, i. e., in $2D$ and $3D$, the topological structures have an intense energy around of the structures. For example, in $2D$, the structure is localized around the center of the kink, meanwhile, in $3D$, the structure is localized around $r=0$. We believe that these magnetic and non-magnetic structures have this energetic behavior due to the type of interaction of the model, as already show in Ref. \cite{Lima}.

To close the discussion of exact solutions, we observe that if the parameter $\lambda\to 0$, the case in $(2+1)D$ will not have topological structures. In this particular case, the model will only accept so-called non-topological solutions. On the other hand, in $(1+1)D$, the model would only admit trivial solutions.

Within the perturbative context, we presented the cuscuton action in a manner similar to the nonlinear sigma model, with the Lagrange multiplier field $\Sigma=e^{\sigma}$ introduced to deal with the square root term, is apparently much more promising. Within this approach, we formulated Feynman diagrams methodology, found the effective propagator of the Lagrange multiplier field, and classified the possible divergences. We found that their structure is restricted. Therefore, we succeeded to develop a perturbative formalism for the cuscuton. Certainly, more its aspects must be studied. Moreover, perhaps other perturbatively consistent theories reducing to the cuscuton model in a certain limit or after integrating over some fields, can be introduced. Searching for such theories is certainly an interesting task.
	
It is interesting to note that the mechanisms discussed in Ref. \cite{BMP}	do not allow for constructing the supersymmetric extension of the cuscuton. However, within the second perturbatively consistent approach, the cuscuton action has been rewritten in a form similar to the nonlinear sigma model whose supersymmetric extension is well known \cite{sig}. Therefore, the possibility of constructing supersymmetric extension of the cuscuton still must be discussed. 

\section*{Acknowledgments}

The authors thank the Conselho Nacional de Desenvolvimento Cient\'{i}fico e Tecnol\'{o}gico (CNPq), Grant No. 301562/2019-9 (A. Yu. Petrov), and Grant No. 308638/2015-8 (CASA), and Coordena\c{c}\~{a}o do Pessoal de N\'{i}vel Superior (CAPES) for financial support. A. Yu. P. is grateful to N. Afshordi for important discussions.

\end{document}